\documentstyle [twocolumn,prb,aps,floats,graphicx] {revtex}

\begin{document}
\draft
\begin{title}
{Trial wave functions with long-range Coulomb correlations for\\
two-dimensional $N$-electron systems in high magnetic fields}
\end{title} 
\author{Constantine Yannouleas and Uzi Landman} 
\address{
School of Physics, Georgia Institute of Technology,
Atlanta, Georgia 30332-0430 }
\date{Phys. Rev. B {\bf 66}, 115315 (2002)}
\maketitle
\begin{abstract}
A new class of analytic wave functions is derived for two dimensional 
$N$-electron $(2 \leq N < \infty)$ systems in high magnetic fields.
These functions are constructed through breaking (at the Hartree-Fock
level) and subsequent restoration (via post-Hartree-Fock methods) of the 
circular symmetry. They are suitable for describing long-range Coulomb 
correlations, while the Laughlin and composite-fermion 
functions describe Jastrow correlations associated with a
short-range repulsion. Underlying our approach is a 
collectively-rotating-electron-molecule picture, yielding for all
$N$ an oscillatory radial electron density that extends throughout the
system.
\end{abstract}
\narrowtext

\section{Introduction}

Two-dimensional (2D) few-electron systems in strong magnetic fields have been 
the focus of extensive theoretical investigations in the last twenty years.
\cite{lau1,lau2,jai2,jai1,gir,haw,haw2,mac,rua,sek,mak,koo,yl1,yl4}
Many of these studies have used the Jastrow-Laughlin \cite{lau2} (JL)
and composite-fermion \cite{jai2} (CF) wave functions, where the dynamics
of electrons in extended fractional quantum Hall (FQH) systems is governed
by the so-called Jastrow correlations. It was shown \cite{hal}
that the JL functions are exact eigenstates of the $N$-electron problem under 
high-magnetic fields for a special short-range interparticle repulsion. 
However, based on close-to-unity overlaps with exact numerical solutions 
\cite{lau1,gir,haw,haw2,mac,rua,sek,mak} 
of the Coulomb problem for few-electron systems (with $N \leq 8$), 
it is believed \cite{lau2,jai2,jai1,hal} that the 
JL/CF functions should not differ significantly from the exact Coulombic 
solutions.

Recent experiments \cite{chan} on electron tunneling into the edges of a
FQH system have found a current-voltage power law behavior, 
$I \propto V^\alpha$, with values for the exponent $\alpha$ that are in
conflict with the universal prediction $\alpha=1/\nu$ derived from Jastrow 
correlations. These findings motivated \cite{tsi0,tsi} detailed exact 
diagonalization studies of FQH systems at the filling factor $\nu=1/3$ with 
up to $N=12$ electrons. These latest studies revealed that the long-range 
Coulomb correlations lead to the formation of stripe-like oscillations in 
the radial electron densities (ED's) which are responsible \cite{tsi0} for 
the observed unexpected behavior of the current-voltage power law. Most 
importantly, the JL functions fail \cite{tsi} to capture these ED 
oscillations, {\it in spite of having overlaps with the exact wave functions 
that are very close to unity\/}.

For the $N$-electron problem in strong magnetic fields and in the disk
geometry (case of quantum dots, QD's), we use in this paper a 
{\it microscopic\/} many-body approach to
derive analytic wave functions that capture the long-range correlations
of the Coulomb repulsion. To obtain analytic results, we specifically 
consider the limit when the confining potential can be neglected compared to 
the confinement induced by the magnetic field.

Underlying our approach is a physical picture of a collectively rotating 
electron molecule (REM) and the synthesis of the states of the system consists
of two steps: First the 
breaking of the rotational symmetry at the level of the single-determinantal 
{\it unrestricted\/} Hartree-Fock (UHF) approximation yields states 
representing electron molecules (EM's, or finite  crystallites). Subsequently 
the rotation of the electron molecule is described
through restoration of the circular symmetry via post Hartree-Fock methods, 
and in particular Projection Techniques \cite{rs} (PT's). Naturally, the 
restoration of symmetry goes beyond the mean-field and yields 
multi-determinantal wave functions. In contrast to the JL/CF functions, our 
analytic functions (applicable for any $N$ and fractional filling)
yield oscillatory radial ED's in agreement with the exact solutions of
the $N$-electron Coulombic system (see below).

\section{METHOD AND TRIAL WAVE FUNCTIONS}

In general, the symmetry-broken UHF orbitals are 
determined numerically. \cite{yl1,yl4,yl5,yl3} 
However, in the case of an infinite 2D electron gas in strong magnetic fields,
it has been found \cite {mz} that such UHF
orbitals can be approximated by analytic Gaussian functions 
centered at different positions $Z_j \equiv X_j+ \imath Y_j$ and forming an 
hexagonal Wigner crystal (each Gaussian representing a localized electron). 
Such displaced Gaussians are written as (here and in the following $\imath 
\equiv \sqrt{-1}$)
\begin{eqnarray}
u(z, &&Z_j) = (1/\sqrt{\pi}) \nonumber \\
&& \times \exp[-|z-Z_j|^2/2] \exp[-\imath (xY_j+yX_j)]~,
\label{gaus}
\end{eqnarray}
where the phase factor is due to the gauge invariance. $z \equiv x-\imath y$,
and all lengths are in dimensionless units of ${l_B}\sqrt{2}$ with 
the magnetic length being $l_B=\sqrt{\hbar c/eB}$. 

In the case of a Coulombic finite $N$-electron system, it has been found 
\cite{koo,yl1} that the UHF orbitals arrange themselves in concentric rings 
forming EM's (referred to also as Wigner molecules, WM's). \cite{note2} The 
UHF results for the ring arrangements are in agreement with the molecular 
structures obtained via the conditional probability distributions (CPD's, 
which can be extracted from exact numerical wave functions
\cite{sek,mak,yl6}), as well as with those obtained \cite{bed} for the 
equilibrium configurations of classical point charges in a 2D harmonic trap.
\cite{bre}

For an $N$-particle system, the electrons are situated at the apexes of 
$r$ concentric regular polygons. The ensuing multi-ring structure is
denoted by $(n_1,n_2,...,n_r)$ with $\sum_{q=1}^r n_q=N$.
The position of the $j$-th electron on the $q$-th ring is given by
\begin{equation}
Z_j^q={\widetilde Z}_q \exp[\imath 2\pi (1-j)/n_q],\;\;  1 \leq j \leq n_q~.
\label{zj1}
\end{equation}

We expand now the displaced Gaussian 
(\ref{gaus}) over the Darwin-Fock single-particle states. Due to the high 
magnetic field, only the single-particle states,
\begin{equation}
\psi_l(z) = \frac{z^l}{\sqrt{\pi l!}} \exp(-zz^*/2)~,
\label{lll}
\end{equation}
of the lowest Landau level are needed (the angular
momentum of this state is $-l$ due to the definition $z \equiv x-\imath y$)
Then a straightforward calculation \cite{mik} yields
\begin{equation}
u(z,Z)=\sum_{l=0}^{\infty} C_l(Z) \psi_l(z)~,
\label{uexp}
\end{equation}
with $C_l(Z)=(Z^*)^l \exp(-ZZ^*/2)/\sqrt{l!}$ for $Z \neq 0$. Naturally,
$C_0(0)=1$ and $C_{l>0}(0)=0$.

Since electrons in strong magnetic fields are fully polarized, only the space 
part of the many-body wave functions needs to be considered. \cite{note3}
The symmetry-broken UHF determinant, $\Psi_{\text{UHF}}^N$, describing the WM 
is constructed out of the localized wave functions $u(z,Z_j^q)$.
Using (\ref{uexp}) one finds the following expansion (within a 
proportionality constant)
\begin{eqnarray}
\Psi_{\text{UHF}}^N =&& \sum_{l_1=0,...,l_N=0}^{\infty} 
\frac{ C_{l_1}(Z_1)C_{l_2}(Z_2) \cdot \cdot \cdot C_{l_N}(Z_N) }
{\sqrt{l_1! l_2! \cdot \cdot \cdot l_N!} }  \nonumber \\
&& \times \; D(l_1,l_2,...,l_N) \exp(-\sum_{i=1}^N z_i z_i^*/2)~,
\label{uhfde}
\end{eqnarray}
where $D(l_1,l_2,...,l_N) 
\equiv {\text{det}}[z_1^{l_1},z_2^{l_2}, \cdot \cdot \cdot, z_N^{l_N}]$. The 
$Z_k$'s (with $1 \leq k \leq N$) in Eq.\ (5) are the $Z_j^q$'s of Eq.\ (2), 
but relabeled.

The UHF determinant $\Psi_{\text{UHF}}^N$ breaks the
rotational symmetry and thus it is is not an eigenstate of the total angular 
momentum $\hbar \hat{L}=\hbar \sum_{i=1}^N \hat{l}_i$. However, one can 
{\it restore\/}  \cite{yl3,rs} the rotational symmetry by applying onto 
$\Psi_{\text{UHF}}^N$ the following projection operator \cite{note9}
\begin{equation}
{\cal O}_L \equiv \prod_{q=1}^r ~{\cal P}_{L_q}~,
\label{genp}
\end{equation}
\noindent
with
\begin{equation}
2 \pi {\cal P}_{L_q} \equiv \int_0^{2 \pi}
d\gamma_q \exp[\imath \gamma_q (\hat{L}_q-L_q)]~,
\label{amp}
\end{equation}
where $\hbar \hat{L}_q=\hbar \sum_{i=i_q+1}^{i_q+n_q} \hat{l}_i$ and
$\hbar L_q=\hbar \sum_{i=i_q+1}^{i_q+n_q} l_i$ with 
$i_q = \sum_{s=1}^{q-1} n_s$ $(i_1=0)$ are partial angular momenta 
operators and values, respectively, associated with the $q$th ring, and 
$\hbar L = \hbar \sum_{q=1}^r L_q$ are the eigenvalues of the total angular 
momentum. 

When applied onto $\Psi_{\text{UHF}}^N$, the
projection operator ${\cal O}_L$ acts as a product of Kronecker deltas: from 
the unrestricted sum (\ref{uhfde}), it picks up only those terms having 
a given total angular momentum $L$ and a specific ordered partition
of it into partial angular momenta associated with the concentric rings, 
i.e., $\hbar L = \hbar \sum_{q=1}^r L_q$. The final analytic expression 
depends on the specific ring arrangement $(n_1,n_2,...,n_r)$. For lack of 
space, we will present here explicitly only the simplest nontrivial 
arrangement, i.e., $(n_1,n_2)$, with more complex [or simpler ones, i.e.,
(0,N) and $(1,N-1)$] obtained via straightforward extensions.

For specific electron locations (\ref{zj1}) associated with the $(n_1,n_2)$ 
WM, one derives \cite{note4} the following symmetry-preserving, many-body 
correlated wave functions (within a proportionality constant),
\begin{eqnarray}
\Phi_{L_1,L_2} && (n_1,n_2;[z]) = \nonumber \\
&& \sum^{l_1 + \cdot \cdot \cdot +l_{n_1}=L_1,\;
l_{n_1+1} + \cdot \cdot \cdot +l_{N}=L_2}%
_{0 \leq l_1<l_2< \cdot \cdot \cdot <l_N}
\left( \prod_{i=1}^N l_i! \right)^{-1}  \nonumber \\
&& \times \left( \prod_{1 \leq i < j \leq n_1} 
\sin \left[\frac{\pi}{n_1}(l_i-l_j)\right] \right) \nonumber \\ 
&& \times \left( \prod_{n_1+1 \leq i < j \leq N} 
\sin \left[\frac{\pi}{n_2}(l_i-l_j)\right] \right)  \nonumber \\
&& \times \; D(l_1,l_2,...,l_N)
\exp(-\sum_{i=1}^N z_i z_i^*/2)~.
\label{phi1}
\end{eqnarray}
In deriving (\ref{phi1}), we took into account that for each determinant 
$D(l_1,l_2,...,l_N)$ in the unrestricted expansion (\ref{uhfde})
there are $N!-1$ other determinants generated from it 
through a permutation of the indices $\{l_1,l_2,...,l_N\}$; these determinants
are equal to the original one or differ from it by a sign only.

Generalizations of expression (\ref{phi1}) to structures with a larger
number $r$ of rings involve for each additional $q$-th ring ($ 2 < q \leq r$):
(I) Inclusion of an additional product of sines with arguments containing 
$n_q$; (II) A restriction on the summation of the associated $n_q$ angular 
momenta.

\section{PROPERTIES OF THE REM WAVE FUNCTIONS} 

We call the correlated wave functions [Eq.\ (\ref{phi1})] the 
REM wave functions. Among the properties of the REM functions,
we mention the following:

1) The REM wave functions lie entirely within the Hilbert subspace spanned by 
the lowest Landau level and, after expanding the determinants, \cite{note4} 
they can be written in the form (within a proportionality constant),
\begin{equation}
\Phi^N_L[z] = P^N_L [z] \exp(-\sum_{i=1}^N z_i z^*_i/2)~,
\label{pl}
\end{equation}
where the $P^N_L[z]$'s are order-$L$ homogeneous polynomials of the $z_i$'s.

2) The polynomials $P^N_L[z]$ are divisible by 
\begin{equation}
P^N_V[z] = \prod_{1 \leq i < j \leq N} (z_i-z_j)~,
\label{pv}
\end{equation}
namely $P^N_L[z]=P^N_V[z] Q^N_L[z]$.
This is a consequence of the antisymmetry of $\Phi^N_L[z]$.
$P^N_V[z]$ is the Vandermonde determinant $D(0,1,...,N)$. For the case of
the lowest allowed angular momentum $L_0=N(N-1)/2$ (see below), one
has $P^N_{L_0}[z] = P^N_V[z]$, a property that is shared with the 
Jastrow-Laughlin \cite{lau2} and composite-fermion \cite{jai2} trial 
wave functions.

3) The $P^N_L[z]$'s are translationally invariant functions.

\begin{table}[t]
\caption{The $Q^3_9[z]$ polynomial associated with the EMWF's and the JL
functions (The $Q^N_L[z]$ polynomials are of order $L-L_0$).}
\begin{tabular}{cc}
EMWF & $~~(z_1^3 -3 z_1^2 z_2 + z_2^3 +6 z_1 z_2 z_3 -3 z_2^2 z_3 -3 z_1 z_3^2
        +z_3^3) $ \\
~& $\times (z_1^3 -3 z_1 z_2^2 + z_2^3 +6 z_1 z_2 z_3 -3 z_1^2 z_3 -3 z_2 z_3^2
        +z_3^3)$ \\ \tableline
JL & $(z_1-z_2)^2 (z_1-z_3)^2 (z_2-z_3)^2$ \\
\end{tabular}
\end{table}

4) The coefficients of the determinants [i.e., products of sine functions, see
Eq.\ (\ref{phi1})] dictate that the REM functions are nonzero 
only for special values of the total angular momentum $L$ given 
for a $(n_1,n_2,...,n_r)$ configuration by,
\begin{equation}
L=N(N-1)/2 + \sum_{q=1}^r n_q k_q,\;\;k_q=0,1,2,3,...~
\label{l0n}
\end{equation}
The minimum angular momentum $L_0=N(N-1)/2$ is 
determined by the fact that the $D$ determinants [see Eq.\ (\ref{phi1})] 
vanish if any two of the single-particle angular momenta 
$l_i$ and $l_j$ are equal. For the $(0,N)$ and $(1,N-1)$ rings, the special 
values are given by $L=L_0+Nk$ and $L=L_0+(N-1)k$, respectively. In plots of
the energy vs. the angular momenta, derived from exact-diagonalization studies,
\cite{rua,sek,mak} it has been found that the special $L$ 
values associated with the $(0,N)$ and $(1,N-1)$ rings (appropriate
for $N \leq 7$) exhibit prominent cusps reflecting enhanced stability; as a 
result these $L$ values are often referred to as ``magic angular momenta''.
\cite{note12} We predict that similar magic behavior reflecting enhanced
stability is exhibited by the special $L$ values given by Eq.\ (\ref{l0n}) 
and associated with the general ring arrangement $(n_1,n_2,...,n_r)$. In the 
thermodynamic limit, \cite{lau2,gir} the total $L$ is related to a
fractional filling $\nu=N(N-1)/(2L)$, and thus the angular momenta (\ref{l0n})
of the REM functions correspond to all the $\nu$ associated with the FQHE, 
including the even-denominator ones, i.e., $\nu=$ 1, 3/5, 3/7, 5/7, 2/3, 1/2, 
1/3, etc... 

5) For the case of two electrons $(N=2)$, the REM functions reduce to
the Jastrow-Laughlin form, namely
\begin{equation}
P^2_L[z]=\prod_{1 \leq i < j \leq N} (z_i-z_j)^L~,
\label{p2}
\end{equation}
where $L=1$, 3, 5, ...
However, this is the only case for which there is coincidence between
the REM and the JL wave functions. 
For higher numbers of electrons, $N$, the polynomials $P^N_L[z]$
of the REM functions 
(apart from the lowest-order Vandermonde $P^N_{L_0}[z]$ ones) are quite
different from the corresponding JL or composite-fermion polynomials.
In particular, the familiar factor $\prod_{1 \leq i < j \leq N} 
(z_i-z_j)^{2p}$, with $p$ an integer, \cite{jai2,jai1} (which reflects 
multiple zeroes) does not appear in the REM functions (see, e.g., 
Table I which contrasts the $Q^3_9[z]$ polynomials corresponding to the 
REM and JL functions).

\begin{figure}[t]
\centering\includegraphics[width=7.5cm]{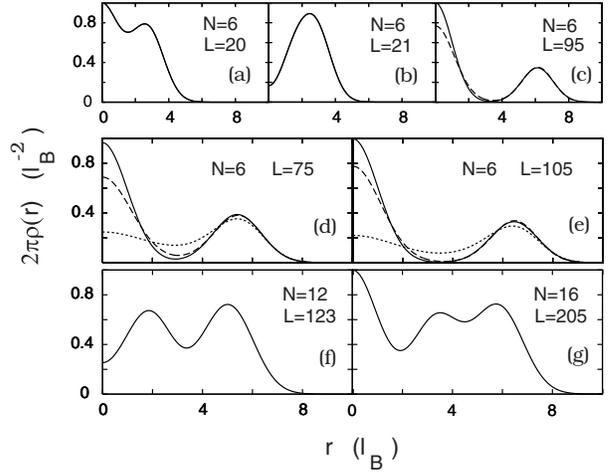}\\ 
\caption{
Radial ED's from EMWF's (solid lines, all frames), exact diagonalization
[dashed lines, (a)$-$(e)], and JL functions [dotted lines, (d) and (e)].
In (a) and (b), the solid and dashed curves are practically
indistinguishable.
}
\end{figure}
6) For the case of three electrons $(N=3)$, after transforming to the
Jacobi coordinates 
$\bar{z}=(z_1+z_2+z_3)/3$, $z_a=(2/3)^{1/2}((z_1+z_2)/2-z_3)$,
$z_b=(z_1-z_2)/\sqrt{2}$ (and dropping the center-of-mass exponential factor),
the REM wave functions can be written as (again within a proportionality
constant),
\begin{eqnarray}
\Phi^3_L[z_a,z_b]=&& [(z_a+ \imath z_b)^{L}-(z_a-\imath z_b)^{L}]
\nonumber \\
&& \times \exp[(-1/2)(z_a z_a^* + z_b z_b^*)]~,
\label{jac}
\end{eqnarray}
with $L=3m$, $m=1$, 2, 3, 4, ... being the total angular momentum. 
Again the wave functions $\Phi^3_L[z_a,z_b]$ are very different from the 
three-electron JL ones; e.g., they are nonvanishing for even $m$
values, unlike the three-electron JL functions. However, the
$\Phi^3_L[z_a,z_b]$'s coincide with the functions $|m,0 \rangle$ derived 
in Ref.\ \onlinecite{lau1}. We remark that, although it was found \cite{lau1}
that these wave functions exhibited behavior expected of fractional
quantum Hall ground states (e.g., areal quantization and incompressibility), 
the generalization of them to a higher number of electrons did not follow;
instead, the REM functions presented here do constitute such a generalization.

\section{OSCILLATORY ELECTRON DENSITIES}

In Fig.\ 1, we display the radial ED's of several REM wave functions and 
compare them to corresponding ED's from exact diagonalizations. The main 
conclusion is that the ED's of the REM functions exhibit \cite{note30} a 
prominent oscillatory behavior in excellent agreement with the exact ED's. 
Such an oscillatory behavior is a 
natural consequence of the underlying ring arrangements. For $N=6$ and $L=20$, 
the underlying structure is a $(1,5)$ arrangement [$L$ is 5 units larger than 
the minimum $L_0=15$, i.e., $n_1=1$, $k_1=0$ and $n_2=5$, $k_2=1$ in Eq.\ 
(\ref{l0n})], and thus the corresponding ED exhibits a maximum at the
origin followed by an outer hump [Fig.\ 1(a)]. For $N=6$ and $L=21$, however, 
the underlying structure is a $(0,6)$ arrangement ($L$ is 6 units larger than 
$L_0$), and thus the ED exhibits a dip at the origin and a single outer hump
[Fig.\ 1(b)]. In the other $N=6$ cases plotted here [Figs.\ 1(c), 1(d), 1(e)],
the difference $L-L_0$ is divisible by 5 and the underlying ring arrangement 
is $(1,5)$; thus there are two humps in the corresponding ED's, the inner one 
portraying the single electron at the origin. The $N=6$, $L=75$ [Fig.\ 1(d) ] 
and $N=6$, $L=105$ [Fig.\ 1(e)] cases correspond to the fractional fillings 
$1/5$ and $1/7$, respectively. For these two cases, we have plotted also the 
ED's associated with the JL functions (dotted lines).
As was found in the $\nu =1/3$ case,
\cite{tsi} the JL functions fail to capture the radial oscillations that are
characteristic of the long-range Coulomb force. Finally, in Fig.\ 1(f) and
Fig.\ 1(g), we present the ED's of the REM functions for $N=12$, $L=123$ and 
$N=16$, $L=205$.
In general, there are as many humps as the number of concentric rings. Indeed 
the ring structures are \cite{koo,bed} (3,9) and (1,5,10) for $N=12$ and 
$N=16$, respectively. \cite{note20} We note that the ED for $N=12$, $L=123$ 
is similar to the exact ED for the same case, \cite{rez} although the latter 
was calculated with an external confinement.

It has been found \cite{koo,bed} that the number of rings $(r)$ increases
as the number of electrons grows. For $\nu \leq 1/3$, this results in 
electron-density oscillations that extend along the whole radius of the QD, 
with no obvious separation into bulk and edge regions.
Currently, the largest number of electrons for which the ring structure has
been determined \cite{bed} is $N=230$ with a concentric-ring arrangement of
(1,6,12,18,23,25,34,37,37,37).

\section{CONCLUSIONS}

We have developed a new class of trial wave functions of 
simple functional form, which accurately describe the physics of electrons in 
QD's under high magnetic fields. In particular, our functions capture the 
long-range correlations of the Coulomb repulsion; unlike the JL 
functions, they yield for all $N$ and fractional fillings $\nu$ an oscillatory
radial electron density in agreement with exact-diagonalization results. 
The electron density oscillations extend throughout the system. The
thematic basis of our approach is built upon the intuitive, but
{\it microscopically\/} supported, picture of collectively rotating electron 
molecules, and the synthesis of the many-body REM wave functions involves 
breaking of the circular symmetry at the UHF level with 
subsequent restoration of this symmetry via a projection technique. While we 
focus here on the strong magnetic-field regime, we note that the REM picture 
unifies the treatment of strongly correlated states of electrons in QD's over 
the whole magnetic-field range. \cite{yl1,yl5,yl3,yl6}
Finally, our REM wave functions, aimed here mainly at treating finite
electron systems (i.e., QD's), can provide in the thermodynamic limit an
alternative interpretation of the FQH effect; namely, the observed hierarchy
of fractional filling factors may be viewed as a signature originating from
the magic angular momenta of rotating electron molecules.

This research is supported by the U.S. D.O.E. (Grant No. FG05-86ER-45234).


\begin{references}
\bibitem{note2}
Depending on the strength of the interelectron repulsion,
WM's can also form at zero magnetic field, see Refs.\ \onlinecite{yl1,yl5,yl3}.
For a CPD/exact-solution study of such WM's at $B=0$, see Ref.\ 
\onlinecite{yl6}.
\bibitem{note3}
In our earlier publications, we have used \cite{yl1,yl4,yl5,yl3} the term
spin-and-space (sS)-UHF to emphasize the breaking of both 
the spin and space symmetries. In this paper, only 
the breaking of the space symmetry is involved.
\bibitem{note9}
The projection operator (\ref{genp}) is a generalization of
the simpler ${\cal P}_L$ operator used for restoring the circular 
symmetry in the simplest (0,N) and (1,N-1) ring arrangements, \cite{yl3} or 
for objects without a ring structure. \cite{rs}
\bibitem{note12}
The magic angular momenta for $N \leq 7$ have been also discussed in the 
context of the geometric configuration \cite{rua,sek} and the Eckart frame
\cite{mak} models. However, unlike our approach, these models do not
afford derivation of analytical forms for the REM wave functions.
\bibitem{note20}
In Eq.\ (\ref{l0n}), the corresponding $k_q$ values are $k_1=1$, $k_2=6$ and
$k_1=0$, $k_2=3$, and $k_3=7$.
\end{references}
\end{document}